\documentclass[aps,prb,twocolumn,showpacs,preprintnumbers,amsmath,amssymb,superscriptaddress,10pt]{revtex4-1}%

\usepackage{graphicx}
\usepackage{amsmath}
\usepackage{color}
\usepackage{bm}
\usepackage {ulem}

\begin{document}
\title{Alternating magnetic anisotropy of Li$_2$(Li$_{1-x}$\textit{T}$_x$)N with \textit{T} = Mn, Fe, Co, and Ni}

\author{A. Jesche}
 \email[]{jesche@ameslab.gov, present address:  Center for Electronic Correlations and Magnetism, Institute of Physics, University of Augsburg,
D-86159 Augsburg, Germany}
\author{L. Ke}
 \affiliation{The Ames Laboratory, Iowa State University, Ames, Iowa, USA}
\author{J. L. Jacobs}
 \affiliation{The Ames Laboratory, Iowa State University, Ames, Iowa, USA}
 \affiliation{Department of Chemistry, Iowa State University, Ames, Iowa, USA}
\author{B. Harmon}
 \affiliation{The Ames Laboratory, Iowa State University, Ames, Iowa, USA}
 \affiliation{Department of Physics and Astronomy, Iowa State University, Ames, USA}
\author{R. S. Houk}
 \affiliation{The Ames Laboratory, Iowa State University, Ames, Iowa, USA}
 \affiliation{Department of Chemistry, Iowa State University, Ames, Iowa, USA}
\author{P. C. Canfield}
 \affiliation{The Ames Laboratory, Iowa State University, Ames, Iowa, USA}
 \affiliation{Department of Physics and Astronomy, Iowa State University, Ames, USA}

\begin{abstract}
Substantial amounts of the transition metals Mn, Fe, Co, and Ni can be substituted for Li in single crystalline Li$_2$(Li$_{1-x}$\textit{T}$_x$)N. 
Isothermal and temperature-dependent magnetization measurements reveal local magnetic moments with magnitudes significantly exceeding the spin-only value.
The additional contributions stem from unquenched orbital moments that lead to rare-earth-like behavior of the magnetic properties.
Accordingly, extremely large magnetic anisotropies have been found.
Most notably, the magnetic anisotropy alternates as easy-plane $\rightarrow$ easy-axis $\rightarrow$ easy-plane $\rightarrow$ easy-axis when progressing from \textit{T} = Mn $\rightarrow$ Fe $\rightarrow$ Co $\rightarrow$ Ni.
This behavior can be understood based on a perturbation approach in an analytical, single-ion model. The calculated magnetic anisotropies show a surprisingly good agreement with the experiment and capture the basic features observed for the different transition metals.  
\end{abstract}

\maketitle
Magnetic anisotropy or magnetic anisotropy energy (MAE) is a fundamental concept in solid state science affecting magnetic data storage, permanent magnets and the investigation of various basic model systems.
In a simple picture, MAE is the energy necessary or reorient magnetic moments in a certain material.
It's value is largely determined by the single-ion anisotropy of the magnetic centers. 
This anisotropy stems from the orbital contribution to the magnetic moment (either directly or indirectly via spin-orbit coupling).
Significant orbital contributions to the magnetic moment and the resulting large MAEs are usually associated with rare-earth elements.
In contrast, the orbital moment in $3d$ transition metals is normally quenched by the crystal electric field. 
Accordingly, the magnetic anisotropy of these elements is often small or non-existent. 

Recently, we have found a remarkable exception to this trend: iron, when substituted in lithium nitride, Li$_2$(Li$_{1-x}$Fe$_x$)N, behaves in many aspects like a rare-earth element\,\cite{Jesche2014b}. 
With an estimated MAE of several hundred Kelvin and, in accordance, an observed coercivity field of more than 11 Tesla this compound exceeds even the largest values observed in rare-earth based permanent magnets. 
Besides iron, other $3d$ transition metal substitutions were synthesized, in polycrystalline form, as early as 1949 by Sachsze and Juza\,\cite{Sachsze1949} and have been subjects of ongoing experimental\,\cite{Gordon2001, Gregory2001a,Niewa2003, Niewa2003b, Schnelle2004, Stoeva2004, Liu2009Li, Muller-Bouvet2014} and theoretical investigations\,\cite{Novak2004, Shunnian2011, Antropov2014}.
It has been found that $T = \rm{Mn, Fe, Co, Ni, and~Cu}$ can be substituted for one of the Li-sites: the two-fold coordinated 1$b$ Wyckoff site. 
Indications for an unusual oxidation state of +1 were also reported\,\cite{Sachsze1949, Niewa2003, Niewa2003b, Schnelle2004, Jesche2014c}.
The transition metals carry a sizable local magnetic moment except for $T = \rm{Cu}$ and highly concentrated $T = \rm{Ni}$\,\cite{Niewa2003, Schnelle2004, Stoeva2004}. 
Due to a lack of large enough single crystals, there has been no direct access to the anisotropy of the physical properties.
Only recently we developed a single crystal growth technique that is based on a lithium-rich flux and is applicable to Li$_2$(Li$_{1-x}$\textit{T}$_x$)N as well as other nitrides and lithium based compounds\,\cite{Jesche2014c}.  

\begin{figure}
\includegraphics[width=8.6 cm]{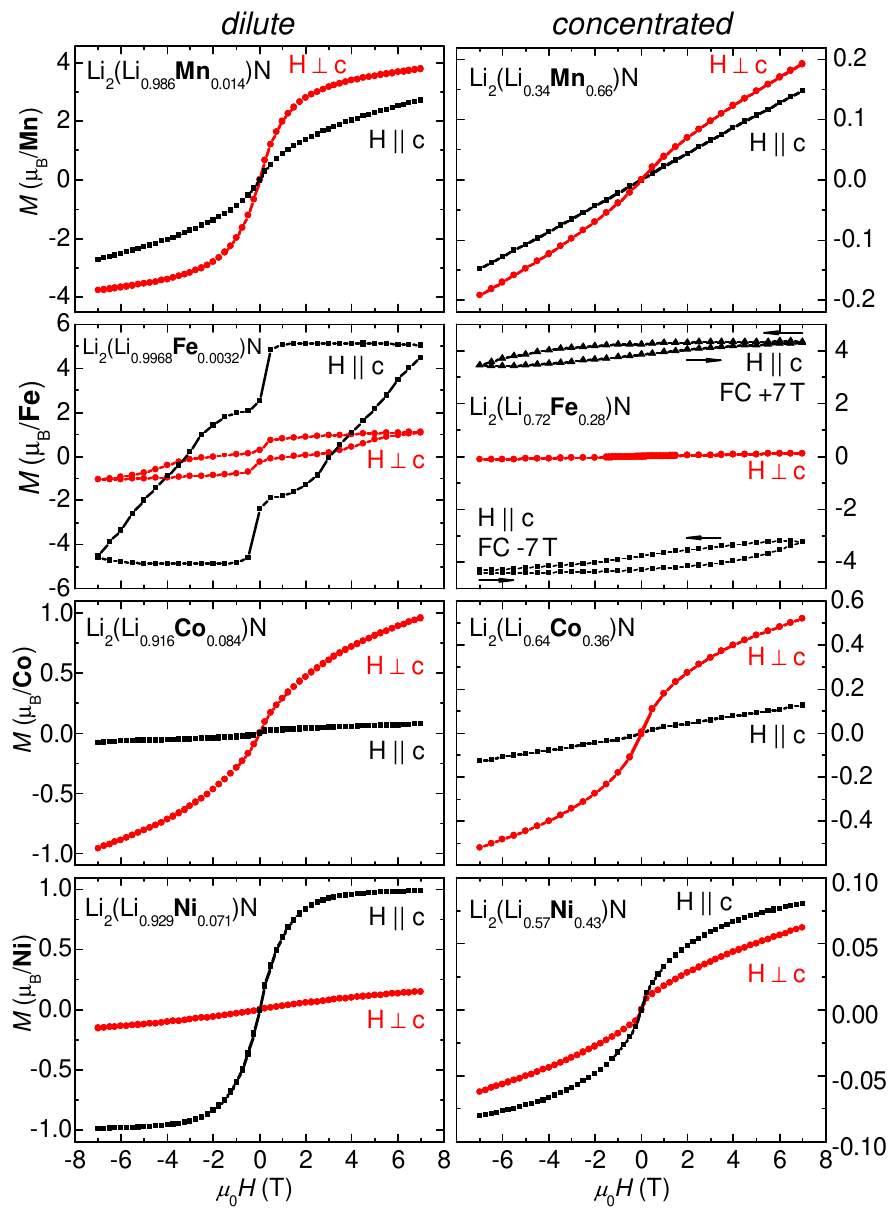}
\caption{(color online) Isothermal magnetization loops at $T = 2$\,K in Bohr magnetons per transition metal ion (FC $\pm7$\,T refers to a field cooled measurement in +7 T and -T, respectively). 
The sign of the magnetic anisotropy changes in an alternating fashion from easy plane to easy axis independent from the concentration of the transition metal. 
}
\label{M-H}
\end{figure}

\begin{figure}
\includegraphics[width=8.6cm]{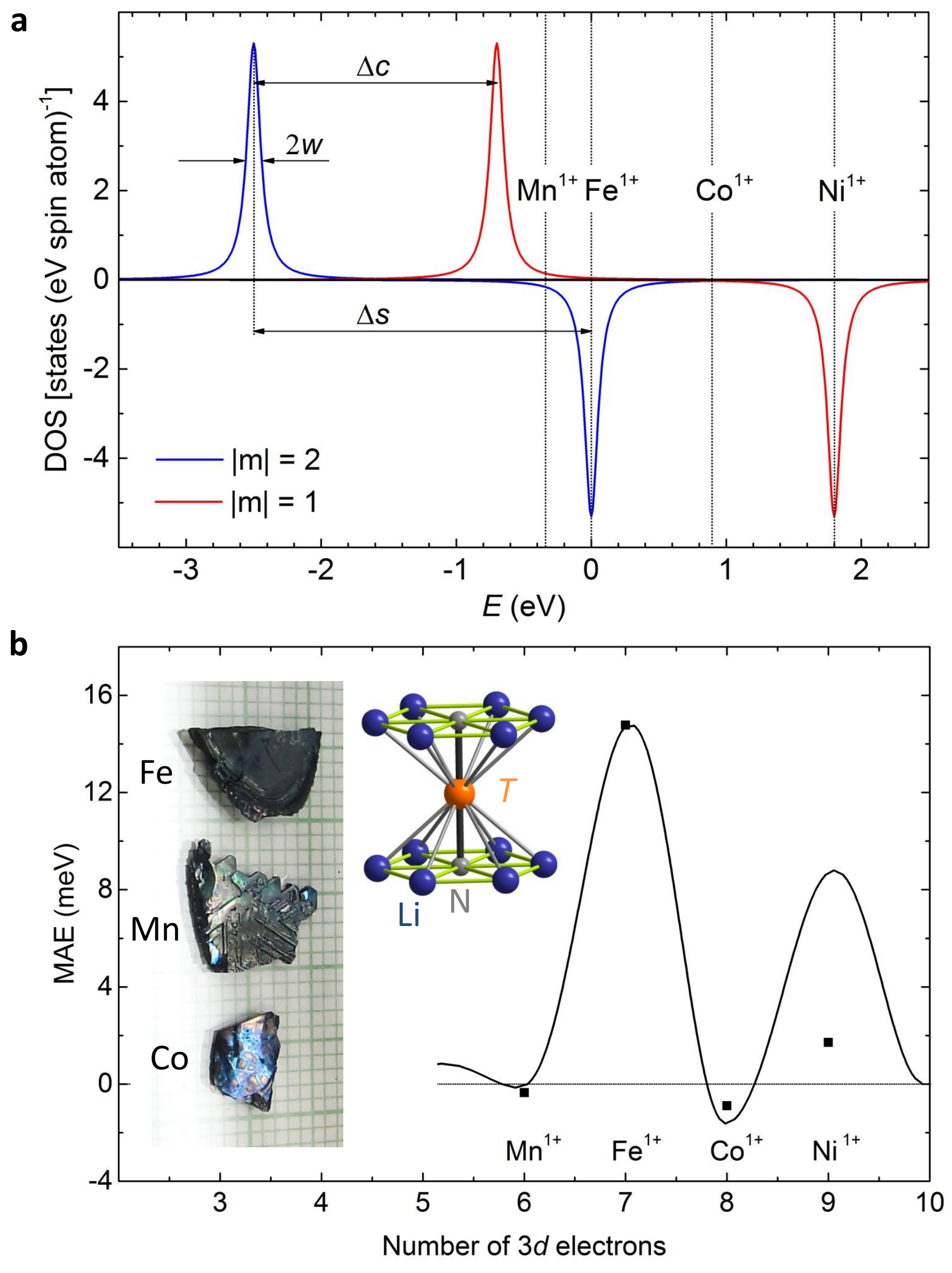}
\caption{(color online) a) Schematic density of states projected on the $3d$ states of isolated Fe atoms.
The MAE in the analytical model is determined by crystal field splitting ($\Delta c$), spin splitting ($\Delta s$), and peak width ($w$).
For $T = \rm{Mn, Co, and~Ni}$ the position of the Fermi level is shifted according to the number of $3d$ electrons (rigid band approximation).
b) The magnetic anisotropy energy (MAE) per isolated \textit{T} atom as a function of the number of $3d$ electrons calculated in the analytical single-ion model is shown by the solid line. The MAE calculated by LSDA is shown for comparison (black squares). 
Inset are three single crystals on a millimeter grid and the local crystal structure of the transition metal with nearest and next-nearest neighbors.
}
\label{theo}
\end{figure}

Here we present the magnetic anisotropy of Li$_2$(Li$_{1-x}$\textit{T}$_x$)N for $T$ = Mn, Co, and Ni and compare the results to our earlier $T = \rm{Fe}$ work. 
Two concentrations, a dilute one and a more concentrated one, were grown, under similar conditions, for the different transition metals. The starting materials were mixed in a molar ratio of Li:$T$:Li$_3$N = 8.97:0.03:1 and 8.7:0.3:1 for dilute and more concentrated samples, respectively. 
The mixtures were packed in a three-cap Ta crucible\,\cite{Canfield2001,Jesche2014c} heated from room temperature to $T = 900^\circ$C over 5\,h, cooled to $T = 750^\circ$C within 1.5\,h, slowly cooled to $T = 500^\circ$C over 60\,h, and finally decanted to separate the single crystals from the excess flux.
A picture of three representative single crystals on a millimeter grid is shown in Fig.\,\ref{theo}b below.
The actual transition metal concentration (as opposed to the nominally melt compositions given above) in the investigated single crystals was determined by chemical analysis using inductively coupled plasma mass spectrometry (ICP-MS)\,\cite{Jesche2014b}.
The ICP-MS instrument was provided by Analytik Jena. 
The deviation from the initial concentration, with respect to nitrogen, differs from one transition metal to the other and also depends on the concentration (see Fig.\,\ref{M-H} for the measured transition metal concentrations). 
However, the obtained concentrations, $x$, clearly reflect the different initial values and allow us to study the \textit{dilute} and \textit{concentrated} regimes. 
Magnetization measurements were performed using a Quantum Design Magnetic Property Measurement System equipped with a 7 Tesla magnet. 
The MAE was calculated analytically in a single-ion model based on second order perturbation theory using the Green's function method. 

Figure\,\ref{M-H} shows the isothermal magnetization in Bohr magnetons per transition metal ion at $T = 2$\,K. The measurements were performed for magnetic field applied parallel and perpendicular to the  hexagonal $c$-axis, shown by black squares and red circles, respectively.
Results obtained for the dilute transition metal substitutions are shown on the left-hand side, more concentrated substitutions are shown to the right.
The largest available field of $\mu_0H = 7$\,T allows for saturation of the magnetization only for Fe and dilute Ni (and comes close to saturation for dilute Mn).
Furthermore, the anisotropy field (crossing point of the $M$-curves for $H \parallel c$ and $H \perp c$) lies well above the largest available fields. 
Therefore, it is not possible to accurately quantify the MAE from our data (for $T \neq \rm{Fe}$). 
However, the alternating change from easy-plane to easy-axis behavior is evident and independent of $x$. Further trends can be recognized: (I) The anisotropy decreases with increasing $x$, except for \textit{T} = Fe. 
(II) The magnetization values that are approached for $\mu_0H = 7$\,T decrease with increasing $x$. 
(III) The anisotropy observed for \textit{T} = Mn is significantly smaller than that observed for the other transition metals (except for concentrated Ni).
(IV) Even though the sign and magnitude of the anisotropy of dilute Fe and Ni appear to be very similar, the large hysteresis found for Fe-substitution is absent for Ni.
(It is also absent for the planar \textit{T} = Mn and Co.) 
Demagnetization fields amount to only a small fraction of the applied magnetic fields and can be neglected.  

\begin{figure}
\includegraphics[width=8.6cm]{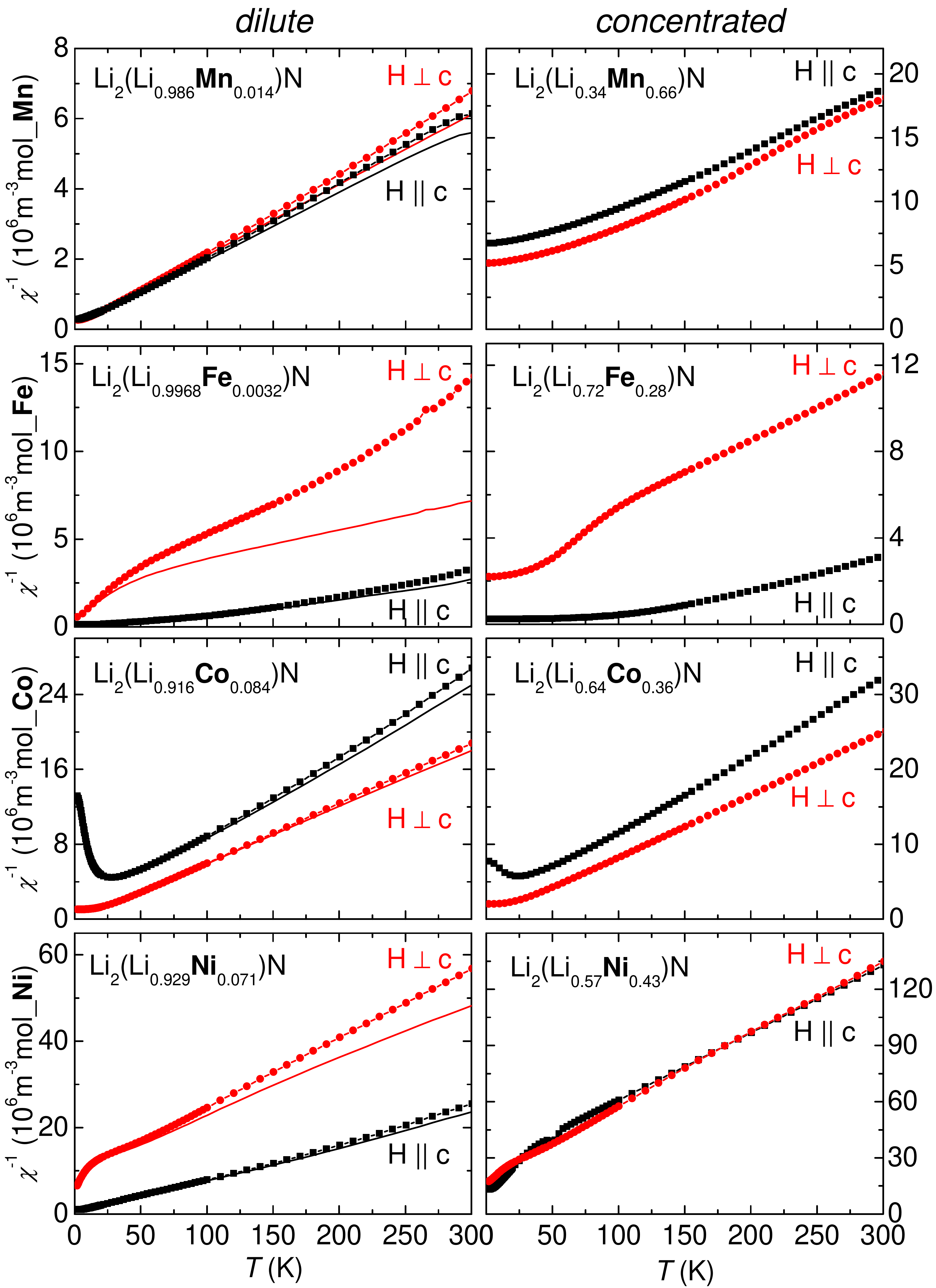}
\caption{(color online) Inverse magnetic susceptibility per mol transition metal as a function of temperature ($\chi^{-1} = H/M$, $\mu_0H = 7$\,T). A pronounced magnetic anisotropy is observed over the whole temperature range except for $T = \rm{Mn}$ and concentrated, $T = \rm{Ni}$ Li$_2$(Li$_{1-x}$\textit{T}$_x$)N. The solid lines given for the dilute case (left panel) show the inverse susceptibility after subtracting the core diagmagnetism of Li$_3$N.}
\label{chi}
\end{figure}

Further information about the orbital moment contributions can be obtained from the effective magnetic moments, determined from the temperature-dependence of the magnetization. 
Figure\,\ref{chi} shows the temperature-dependent magnetic susceptibility, $\chi(T) = M/H$, for both orientations of the applied field. 
Reasonable agreement with Curie-Weiss behavior is observed for $T > 150$\,K. 
The contribution of the core diamagnetism of the Li$_3$N host significantly affects the diluted systems but can be neglected for the larger concentrations shown on the right hand panel of Fig.\,\ref{chi}. 
Subtracting the ionic contributions of Li$^{1+}$\,\cite{Hurd1967} and N$^{3-}$\,\cite{Hohn2009} leads to better agreement with the expected linear behavior of $\chi^{-1}(T)$ over a wide temperature range (solid lines in Fig.\,\ref{chi}). 
The summary of the obtained $\mu_{\rm eff}$ values and a comparison with the spin-only values are given in Table\,\ref{tab_mu} (extracted from the temperature range $150\,\rm{K} < T < 300$\,K).

\begin{table}
\caption{Measured effective magnetic moments per transition metal ion in Li$_2$(Li$_{1-x}$\textit{T}$_x$)N and the spin-only value calculated for $T^{1+}$ in units of Bohr magneton.}
\begin{tabular}{cccccc}
\hline
\hline
    & \multicolumn{2}{c}{dilute~~~} &\multicolumn{2}{c}{concentrated}	&~	\\

\textit{\,T}		~	&	$H \parallel c$	&	~$H \perp c$~ & 	$H \parallel c$	 &	~$H \perp c$~   & ~spin-only      \\
\hline
Mn	&	5.5	&	5.2	&	3.6	&	3.6	&	4.9		\\
Fe	&	6.7	&	$3.7$&	6.5	&	4.6	&	3.8		\\
Co	&	2.6	&	3.1	&	2.5	&	2.7	&	2.8		\\
Ni	&	2.6	&	2.0	&	1.3	&	1.3	&	1.7		\\	
\hline
\hline
\end{tabular}
\label{tab_mu}
\end{table}

In the diluted systems $\mu_{\rm eff}$ significantly exceeds the spin-only value for the easy-axis systems with \textit{T} = Fe and \textit{T} = Ni  in particular for field applied along the easy-axis. 
The diluted and the concentrated system of $T$ = Fe show the largest enhancement of $\mu_{\rm eff}$ when compared to the spin-only value and accordingly the magnetic anisotropy observed in the isothermal magnetization measurements, $M(H)$, is by far the largest among the different transition metals (Fig.\,\ref{M-H}).
In the easy-plane systems $\mu_{\rm eff}$ is only slightly enhanced. For the case of $T$ = Mn this is directly reflected in the rather low magnetic anisotropy observed in $M(H)$. 
Furthermore, the observation of smaller effective moments for concentrated $T$ = Mn when compared with the dilute system goes hand in hand with a further decrease of the anisotropy in $M(H)$.
In a similar fashion, the decrease of the effective moments for concentrated systems of both $T$ = Co and $T$ = Ni when compared to the dilute case is accompanied by a corresponding decrease of the magnetic anisotropy in $M(H)$.

However, for dilute $T$ = Co, a large magnetic anisotropy is apparent in $M(H)$ (Fig.\,\ref{M-H}) even though the effective moments are only slightly enhanced when compared to the spin-only value. 
A more sophisticated analysis is needed to describe the observed behavior.
The input parameters for the employed analytical model are obtained from a simplified representation (Fig.\,\ref{theo}a) of the detailed electronic band structure.
These were calculated earlier in the framework of local spin density approximation (LSDA) for two different supercells which correspond to Fe-concentrations of $x = 0.17$\,\cite{Antropov2014} and $x = 0.50$\,\cite{Klatyk2002, Novak2002, Antropov2014}, respectively.
The calculated densities of states (DOS) projected on the $3d$ states of isolated Fe-atoms turn out to be fairly similar. 
In particular, a sharp peak in the minority spin channel that intersects the Fermi level appears in both cases.
A further dilution of the Fe-concentration is not expected to cause tremendous changes of the projected Fe-DOS, since the Fe-atoms are already well separated.

In our analytical model we consider only the DOS of the four $3d$ states with $m = \pm 1$ and $m = \pm 2$ and model them by Lorentzian-shaped peaks with a half-width $w = 60$\,meV ($m$: orbital quantum number). 
The $m = 0$ states ($3d_{z^2}$) are not included in the modeling since they are well below the Fermi level and have negligible effects on the MAE.
The MAE depends on the crystal field splitting $\Delta c$ (energy difference between $m = \pm 1$ and $m = \pm 2$ states), on the spin splitting $\Delta s$ (energy difference between spin-up and spin-down), and on $w$\,\cite{Ke2015}.
To facilitate the calculation of the MAE for $T = \rm{Mn, Co, and~Ni}$, only the Fermi level is shifted according to the number of $3d$ electrons leaving the band structure unchanged (rigid band approximation). 
Since each degenerate state, and accordingly each Lorentzian in Fig.\,\ref{theo}a, can accommodate two electrons, the Fermi level either intersects the degenerate peaks or sits in the middle, between two PDOS peaks (for an integer electron number). 
For the case of $T = \rm{Mn^{1+}}$, six $3d$ electrons have to be considered. 
Five of these occupy the majority band (upper panel in Fig.\,\ref{theo}a) and the remaining one the $m = 0$ state of the minority band (lower panel in Fig.\,\ref{theo}a, the $m = 0$ state lies well below the Fermi level and is not shown).
Accordingly, the Fermi level for $T = \rm{Mn^{1+}}$ is located between the $ m  = \pm 1$ spin-up and the $ m  = \pm 2$ spin-down states. 
For the case of $T = \rm{Fe^{1+}}$, one more electron has to be placed in the minority band. 
Therefore, one of the two $ m  = \pm 2$ spin-down states has to be occupied. Accordingly, the Fermi level intersects the center of the $m = \pm 2$ spin-down peak. 
Adding one more electron shifts the Fermi level right between the $ m = \pm 2$ and $m = \pm 1$ spin-down states corresponding to $T = \rm{Co^{1+}}$ with eight $3d$ electrons.
For $T = \rm{Ni^{1+}}$, one more $3d$ electron has to be considered and, similar to the case of $T = \rm{Fe^{1+}}$, the  Fermi level intersects the center of the $m = \pm 1$ spin-down peak. 
Furthermore, this allows for the calculation of the MAE as a continuous function of the band filling.
It is worth mentioning that the spin orbit coupling is not included in the schematic band structure. 

The pairwise orbital susceptibility, and therefore the MAE, is now fully determined by $w$ and the energy difference between the Fermi level and the involved orbitals. The later one includes the effect of spin splitting and crystal electric field splitting.
The pair susceptibility is proportional to the term $w/{[}(\epsilon_{\rm F}-\epsilon_m^\sigma)^2 + w^2{]}$, where $\epsilon_{\rm F}$ is the Fermi energy, $\epsilon_m^\sigma$ the DOS peak position with $\sigma$ = \{spin-up, spin-down\}\,\cite{Ke2015}.
For $T = \rm{Fe}$ the $m = \pm 2$ spin-down state is located right at the Fermi level, that is $\epsilon_{\rm F} = \epsilon_2^\downarrow$.   
This leads to a resonance-like enhancement of the pair susceptibility that is only limited by the peak width $w$. 
The MAE for $T = \rm{Fe}$ is therefore dominated by the contribution of the $m = \pm 2$ spin-down state. This result is in full analogy to the LSDA-based calculation, that revealed a splitting of this state caused by spin-orbit coupling which leads to large MAE values\,\cite{Klatyk2002, Novak2002}. For $T = \rm{Mn}$ and Co no such resonance of the pair susceptibility occurs and the resulting MAE, which is determined by the sum over all orbital pairs, turns out to be negative. 

The MAEs calculated in our analytical model are shown in Fig.\,\ref{theo}b as solid line (positive values correspond to an easy-axis, negative ones to an easy-plane system). 
Values obtained by LSDA methods\,\cite{Ke2015} are given for comparison (square data points).
The MAE of Mn, Fe, Co, and Ni correspond to the respective integer number of $3d$ electrons.
Most remarkably, this simplified model is sufficient to capture all basic features of the magnetic anisotropy even though the 'exact' band structure\,\cite{Novak2004,Antropov2014} differs in several details from the schematic representation shown in Fig.\,\ref{theo}a. 
The largest MAE is calculated for $T = \rm{Fe}$. 
Given the simplicity of the model, the calculated value of 15\,meV is in reasonable agreement with our experimental result of 13\,meV for the dilute and 27\,meV for the more concentrated case (estimated from the linearly extrapolated anisotropy field and the measured saturation magnetization) and with the LSDA result. 
The MAE calculated for $T = \rm{Co}$ is significantly smaller and of opposite sign. 
The smallest MAE is calculated for $T = \rm{Mn}$. 
Both, sign and relative magnitude of the calculated MAE fit well to the experimentally observed magnetic anisotropy (Fig.\,\ref{M-H}).
One exception is observed for the $T = \rm{Ni}$ system. The single-ion model gives a MAE of about half of the $T = \rm{Fe}$ value that results mainly from an $\vert m \vert^2$ dependence of the MAE\,\cite{Ke2015}. However, the experiment suggests that the MAE for \textit{concentrated} $T = \rm{Ni}$ is reduced by more than an order of magnitude when compared to $T = \rm{Fe}$.
This discrepancy is mainly caused by an underestimation of the DOS peak width of the $\vert m \vert = 1$ spin-down states that turns out to be larger than the $\vert m \vert = 2$ state (the MAE decreases with increasing band width, roughly following 1/$w$). 
Such a peak-broadening corresponds to an increasing delocalization which is indeed manifested in the decreasing electrical resistivity of Li$_2$(Li$_{1-x}$Ni$_x$)N for $x \gtrsim 0.8$\,\cite{Schnelle2004}.
Furthermore, spin-splitting and crystal field splitting for $T = \rm{Ni}$ are smaller than for $T = \rm{Fe}$\,\cite{Ke2015, Antropov2014}.
Adjusting the schematic band structure accordingly leads to a better agreement between single-ion model and experiment.
There is also good agreement between our experimental results and recent calculations of the MAEs based on LSDA calculations\,\cite{Antropov2014}.

The orbital magnetic moment and the associated large MAE are likely direct consequences of the local symmetry of the transition metal. 
This actual linear, two-fold coordination between the nearest-neighbor nitrogen atoms gives rise to an effective, linear molecule (N-$T$-N). 
And as such, it is not subject to the Jahn-Teller effect\,\cite{Jahn1937} which is driven by lifting the orbital degeneracy. 
Therefore, a quenching of the orbital magnetic moment by a lattice distortion does not take place in Li$_2$(Li$_{1-x}$\textit{T}$_x$)N. 
Within this symmetry the changing of the transition metal gives rise to a dramatic change of both experimental and calculated anisotropies.
It remains to be seen whether this behavior is generic to linear complexes or restricted to the special case of Li$_2$(Li$_{1-x}$\textit{T}$_x$)N.
Further indications for the relevance of a linear arrangement to the formation of orbital magnetic moments and large MAE in $3d$ transition metals can be found in seemingly unrelated systems: Ad-atoms on surfaces\,\cite{Rau2014} (diatomic molecules built from substrate oxygen and adsorbed cobalt) and some linear transition metal complexes\,\cite{Power2012} indeed show significant orbital contributions to the magnetic moment. 

In summary, we found significant orbital contributions to the magnetic moment of the transition metals Mn, Fe, Co, and Ni substituted in Li$_3$N. 
In accordance, large magnetic anisotropies are observed. 
A sharp peak of the DOS which is intersected by the Fermi level gives rise to the uniaxial magnetic anisotropy of Fe and Ni. 
Even though this is not the case for Mn and Co, the latter one does also show a sizable magnetic anisotropy which is, however, of easy-plane type. 
This behavior can be described in an analytical, single-ion model based on only three parameters: crystal field splitting, spin splitting, and peak width of the DOS.
Based on these considerations, it could be possible to identify, or even design, further magnetically ordered transition metal compounds with large orbital magnetic moments and magnetic anisotropy without relying on detailed band structure calculations and excessive computer power. 

\begin{acknowledgments}
The authors want to thank V. Taufour, A. Kreyssig, S. Thimmaiah, W. R. Meier, and R. W. McCallum for fruitful discussions. This work was supported by the U.S. Department of Energy, Office of Basic Energy Science, Division of Materials Sciences and Engineering. The theoretical work was supported by the U.S. Department of Energy - Energy Efficiency and Renewable Energy, Vehicles Technology Office, PEEM program. The research was performed at the Ames Laboratory. Ames Laboratory is operated for the U.S. Department of Energy by Iowa State University under Contract No. DE-AC02-07CH11358.
\end{acknowledgments}

\end{document}